# Large enhancement of sensitivity in NiFe/Pt/IrMn-based planar Hall sensors by modifying interface and sensor architecture

H. Pişkin* and N. Akdoğan

Gebze Technical University, Department of Physics, 41400 Gebze, Kocaeli, Turkey

***Corresponding author:** hpiskin@gtu.edu.tr

**Keywords:** Magnetoresistive sensors, interface effects, planar Hall effect, exchange bias.

## Abstract

The planar Hall sensitivity of obliquely deposited NiFe(10)/Pt($t_{Pt}$) /IrMn(8)/Pt(3) (nm) trilayer structures has been investigated by introducing interfacial modification and altering sensor geometry. The peak-to-peak PHE voltage ($\Delta V_{PHE}$) and AMR ratio of the sensors exhibit an oscillatory increase as a function of Pt thickness. This behaviour was attributed to the strong electron spin-orbit scattering at the NiFe/Pt interface of the trilayers. The temperature-dependent PHE signal profiles reveal that the Pt-inserted PHE sensors are stable even at 390 K with a high signal-to-noise ratio and an increased sensitivity due to reduction of exchange bias. In order to further increase the sensitivity, we have fabricated PHE sensors for a fixed Pt thickness of 8 Å by using sensor architectures of a cross, tilted-cross, one-ring and five-ring junctions. We have obtained a sensitivity of 3.82 μV/Oe.mA for the cross junction, while it considerably increased to 298.5 μV/Oe.mA for five-ring sensor geometry. The real-time voltage profile of the PHE sensors demonstrate that the sensor states are very stable under various magnetic fields and sensor output voltages turn back to their initial offset values. This provides a great potential for the NiFe/Pt/IrMn-based planar Hall sensors in many sensing applications.

## 1. Introduction

Among the magnetoresistive sensors, the planar Hall effect (PHE) based systems have attracted much attention due to their high signal to noise ratio, linear response at low magnetic fields, small offset voltage in the absence of magnetic field, low power consumption and insensitiveness to small thermal fluctuations[1]. With these unique properties, the PHE sensors have been investigated for their potential applications in micro-crack detectors[2], high accuracy electronic micro-compasses[3], magnetoresistive biosensors[4], on-chip magnetometers[5], heart beat counters and blood pressure detectors (tactile sensing)[6]. Recent efforts in this field have been concentrated to increase the sensor sensitivity and to integrate them with lab-on-chip platforms.

The PHE sensor signal have a quite linear zone that describes the sensor working range. The slope of this linear zone defines the sensor's magnetic field sensitivity which can be manipulated by the geometry and structure of the sensor. Up to now, several sensor geometries such as cross junction[7], tilted cross junction[8], Wheatstone bridges in ring [9,10] and diamond[11,12] shapes have been studied. Furthermore, each layer of the sensor structure has influence on the sensor signal by creating shunt current and/or modifying the anisotropy energy of the sensing layer. In the literature, Ni thin film[13], NiFe/IrMn bilayer [14,15][15] NiFe/X/IrMn trilayer (X: Cu[16], Au[17]) and NiFe/Cu/NiFe/IrMn spin-valve[18,19] systems have extensively been studied. Among these structures, the trilayer system provides the highest sensor sensitivity due to its reduced exchange bias compared to the bilayers and lower shunt current compared to the spin-valves structures [16].

In addition, the type and thickness of the spacer layer in trilayer structures strongly affect the PHE sensor sensitivity. Except one group used Au[17], in all trilayer-based PHE sensors studies thin Cu layer is used as a spacer[12,16,20–22]. Although the Cu spacer magnetically softens the NiFe sensing layer[23][24], it results in a decreased sensor output voltage due to its low resistivity[16]. This negative effect of Cu on the sensor sensitivity becomes dominant especially for thicker spacer layers. In order to avoid these effects, a spacer material with a higher resistivity must be used by optimizing its thickness for PHE sensor applications.

In this study, considering that Pt has strong spin–orbit coupling and much higher resistivity compared to the Cu spacer layer, we have investigated the PHE sensor sensitivity as a function of Pt spacer layer thickness in NiFe/Pt/IrMn trilayers. The sensor elements were deposited by magnetron sputtering and photolithography methods. Magnetization and transport measurements reveal that the sensor sensitivity increases as a function of Pt spacer thickness due to decreased exchange bias in trilayers. Notably, we have observed an oscillatory enhancement of PHE sensor output voltage and AMR ratio by inserting thicker Pt layers. Temperature-dependent measurements indicate that the produced PHE sensors are very stable even at 390 K. Furthermore, we have prepared different sensor geometries to increase the sensor sensitivity for a fixed sensor structure. We have obtained 298.5 μV/Oe.mA sensor sensitivity for five-ring geometry. We have also carried out timescan experiments under different magnetic fields in order to investigate the magnetic bead detection capability and the stability of the produced PHE sensors.

## 2. Sensor fabrication

The PHE sensors with architectures of Hall bar, cross, tilted-cross, one-ring and five-ring were fabricated by the lift-off of a trilayer structure of NiFe(10)/Pt($t_{Pt}$) /IrMn(8)/Pt(3) (nm) deposited on Si substrates by magnetron sputtering. The nominal thickness of the Pt spacer layer was varied from 0 nm to 1 nm with a step of 1 Å. $Ni_{80}Fe_{20}$, Pt and $Ir_{22}Mn_{78}$ layers were sequentially deposited at room temperature by using 10W DC, 10W DC and 20W DC powers, respectively. Since the PHE sensors were fabricated in zero applied field, the easy axis of the magnetization was formed parallel to the current axis by growth-induced magnetic anisotropy.

Fig. 1(a) shows a picture of the sample on a printed circuit board (PCB) with a Hall bar geometry and simultaneously grown continuous film. The six-terminal Hall bars have a 25 μm×25 μm active area. Fig. 1(b) presents microscopic images of cross, tilted-cross, one-ring and five-ring junctions. The width of the cross and tilted-cross junctions was 5 μm and the tilt angle was 45°. The outer radii of the ring sensors were kept 150 μm with a ring width of 5μm.

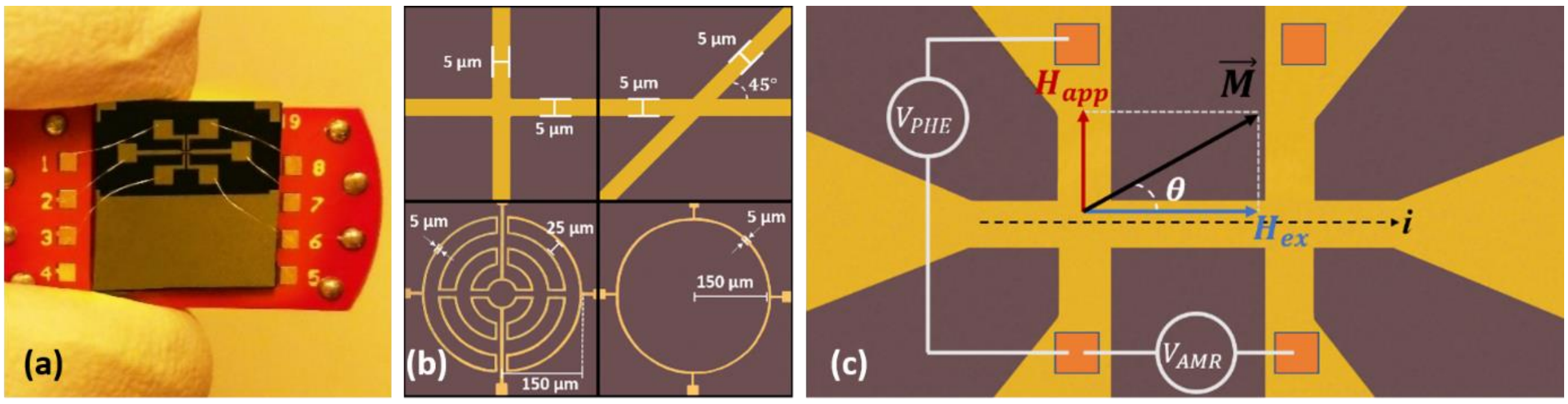

Figure 1. (a) A photograph of the sample mounted on a PCB which contains both Hall bar and continuous film parts. (b) Microscopic images of the PHE sensors with the architectures of cross, tilted-cross, one-ring and five-rings. (c) Schematic representation of the current, magnetization and applied field directions, easy axis of the sensors, exchange bias field and the angle (θ) between the magnetization and easy axis.

## 3. Experimental results

Magnetic hysteresis curves, exchange bias and coercive field of the sensors were measured by using a magneto-optical Kerr effect (MOKE) setup which is equipped with a flow cryostat for temperature-dependent experiments[25]. The easy axis hysteresis loops of three samples were shown in Fig. 2(a). When the Pt spacer layer were not inserted into the NiFe/IrMn interface, the values of coercive field ($H_C$) and exchange bias ($H_{EB}$) are very large due to strong pinning. The insertion of Pt weakens the pinning and shifts the hysteresis loops to lower negative fields. This systematic decrease in both $H_{EB}$ and $H_C$ as a function Pt spacer thickness can be seen in Fig. 2(b). It is also important to note that exchange bias does not disappear in trilayer system even with a Pt spacer thickness of 1 nm.

We also measured PHE and AMR voltage profiles of the sensors using a Keithley 2002 multimeter. A sensor current of 5 mA was applied for Hall bar geometries, whereas 1mA was applied for all other sensor architectures using a Keithley 2400 sourcemeter. The measurement geometry is presented in Fig. 1(c). The PHE sensitivity of the sensors were calculated from the linear region of the PHE voltage profiles and given in Fig. 2(c) as a function Pt spacer thickness. Fig. 2(c) indicates that the PHE sensitivity of the Pt inserted trilayers were increased from 0.86 µV/Oe.mA to 5.32 µV/Oe.mA due to decrease in $H_{EB}$.

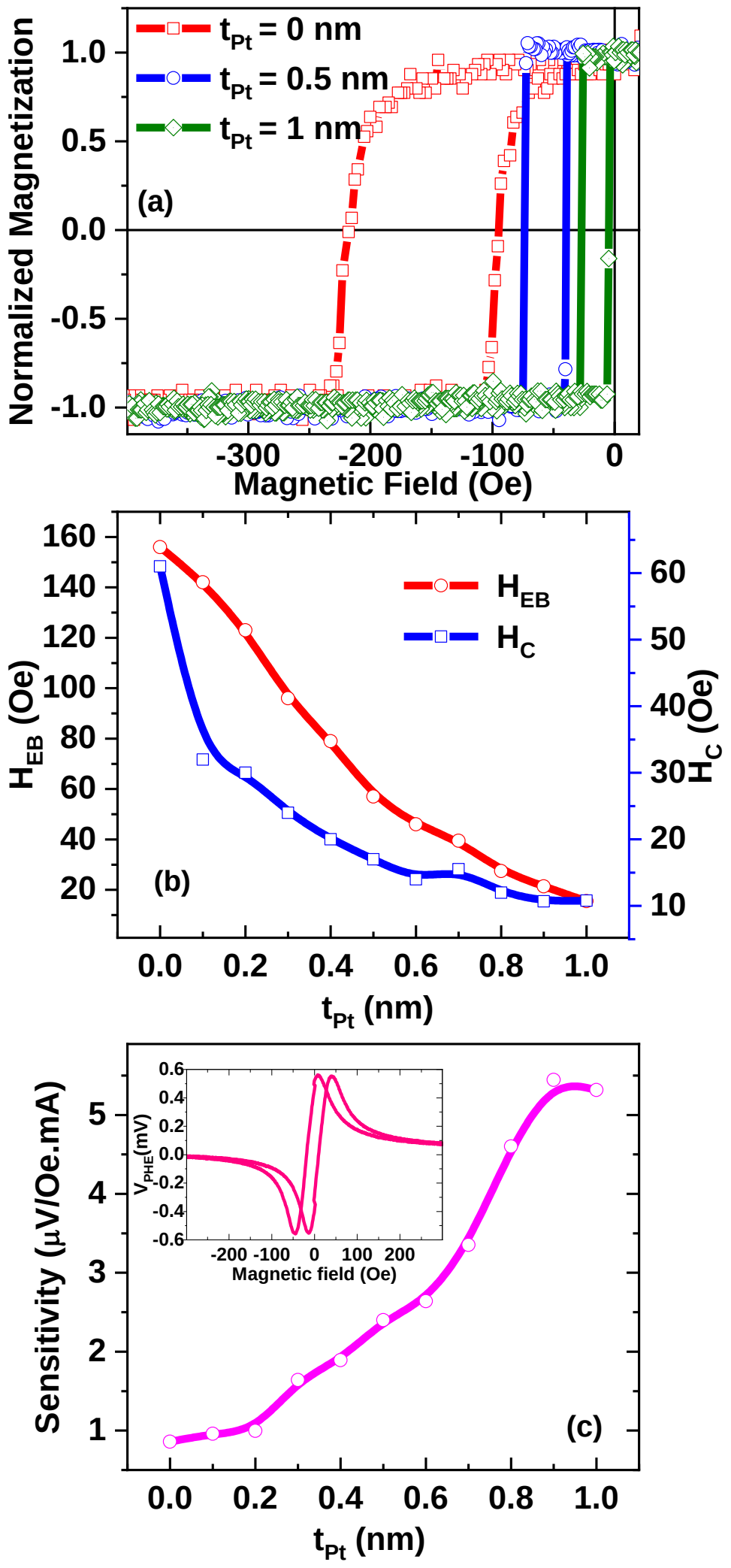

Figure 2. (a) Easy axis hysteresis loops of trilayers taken by MOKE setup for certain thicknesses of Pt spacer layer. (b) $H_C$ and $H_{EB}$ values of the samples as a function of Pt thickness. (c) Pt thickness dependence of the PHE sensor sensitivities calculated from the linear regions of PHE voltage profiles. Inset shows the PHE measurement of the Hall bar with a Pt spacer of 1 nm. Solid lines are guide for the eye.

We have also plotted peak-to-peak PHE voltage ($\Delta V_{PHE}$) and AMR ratio of the sensors in Fig. 3 as a function of Pt spacer thickness. In principle, the addition of each conductive layer to the sensor structure reduces the current that passing through the sensing layer by creating a shunting effect. Thus, the $\Delta V_{PHE}$ and AMR ratio of trilayers were expected to decrease as a function Pt thickness. On the contrary, we have observed an oscillatory increase of $\Delta V_{PHE}$ and AMR ratio in trilayers by addition of Pt spacer. This is attributed to the enhancement of resistivity difference ($\Delta\rho$=$\rho_{\parallel} - \rho_{\perp}$) in the sensing layer due to strong electron spin-orbit scattering at the NiFe/Pt interfaces of the trilayers. The similar behaviour of the resistivity difference at the NiFe/Pt interface were also reported in Ref. [26]. However, the oscillatory behaviour of this enhancement observed in the present work requires further studies and analysis to understand its origin.

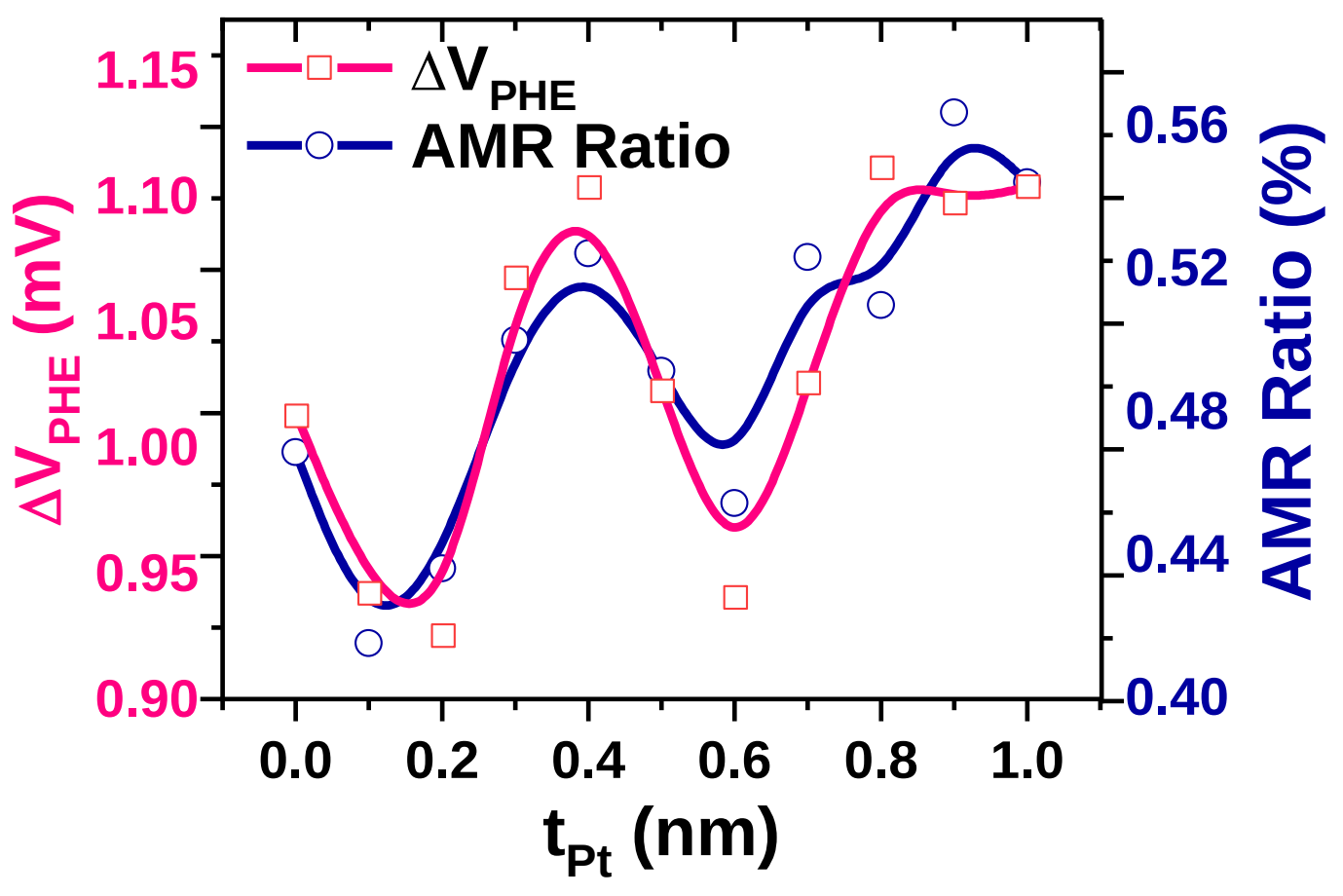

Figure 3. The peak-to-peak PHE voltage ($\Delta V_{PHE}$) and AMR ratio of the sensors as a function of Pt spacer thickness. Solid lines are guide for the eye.

In order to investigate the high temperature characteristics of the PHE sensors, we have performed temperature-dependent measurements for a fixed spacer layer thickness of 3 Å. Fig. 4 shows the PHE sensitivity of the sensor as a function of temperature scanned from 300 K to 390 K with a step of 10 K. It is clear from the data that the sensitivity was enhanced from 2.0 μV/Oe.mA to 5.1 μV/Oe.mA due to the reduced exchange bias as a function of temperature. It is worthy of note that the sensor is still very stable even at 390 K with a good sensitivity and high signal-to-noise ratio. The PHE voltage profiles given in the inset of Fig. 4 indicate that the peak-to-peak PHE voltage value ($\Delta V_{PHE}$) decreases at high temperatures due to decrease of the resistivity difference ($\rho_{\parallel} - \rho_{\perp}$). However, the maximum of the PHE voltage shifts to the lower magnetic fields as a function of temperature. This shift results in an increase in the slope of the PHE signals and it is attributed to the decrease of the $H_{EB}$ at high temperatures.

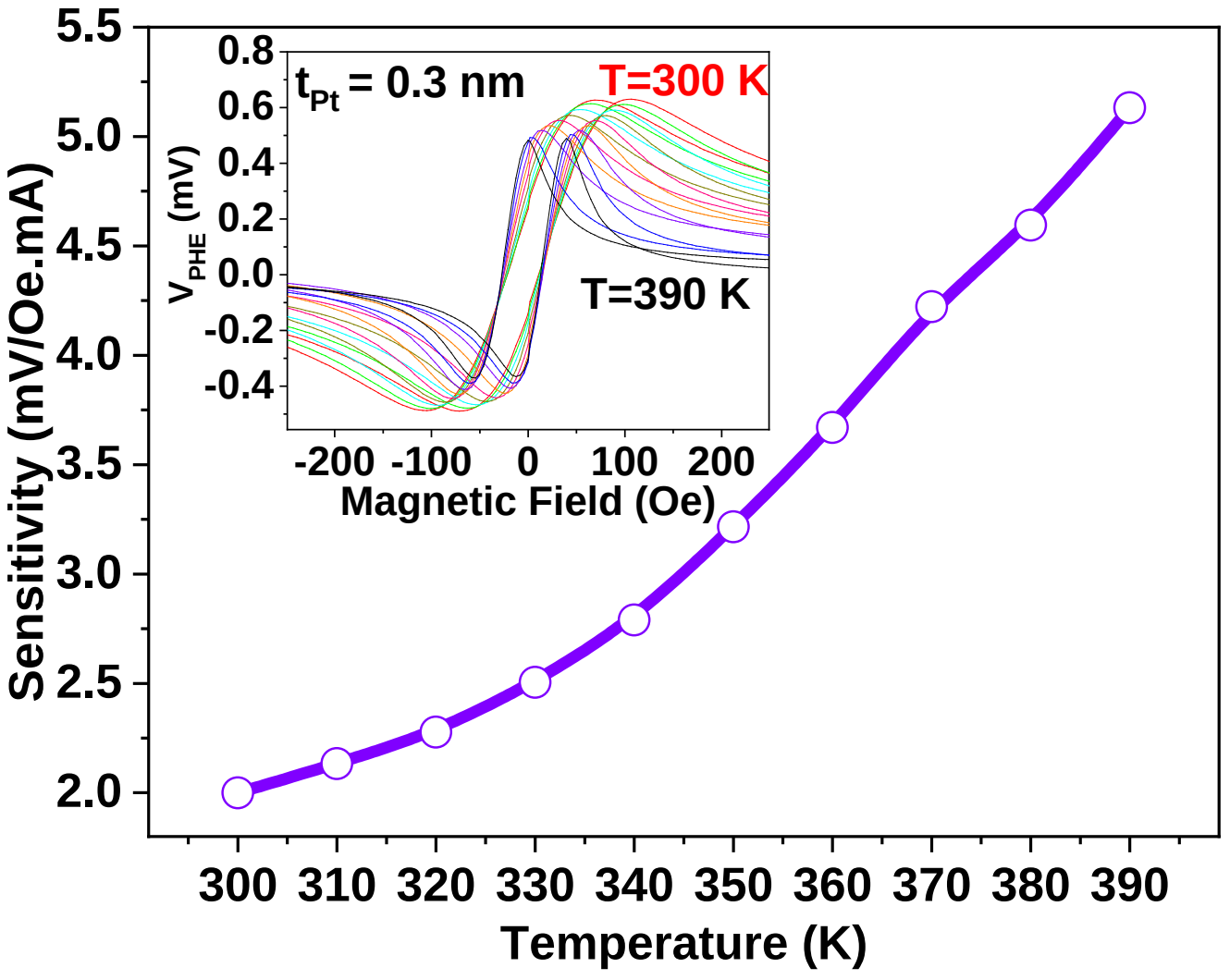

Figure 4. Temperature-dependence of the PHE sensor sensitivity for a trilayer structure with a Pt spacer of 3 Å. Solid line is a guide for the eye. Inset presents PHE voltage profiles taken at different temperatures from 300 K to 390 K.

Furthermore, we have investigated the geometry dependence of the sensitivity in Pt inserted PHE sensors for a fixed Pt thickness of 8 Å. The sensor output signals of four different geometries were given in Fig. 5. We have observed a sensitivity of 3.82 μV/Oe.mA for the cross junction, and it slightly increased to 5.80 μV/Oe.mA for the tilted-cross architecture. However, we have recorded huge increase in the sensitivities of the ring junctions. The obtained sensor sensitivities were 96.08 μV/Oe.mA and 298.46 μV/Oe.mA for one-ring and five-ring junctions, respectively. Indeed, it was expected to observe higher sensitivities in tilted-cross and ring junctions compared to the cross junction for a fixed sensor structure. Because, the AMR and Wheatstone bridge contributions were added to the conventional PHE voltage in the tilted-cross and ring geometries [8,10]. These contributions did not change the position of the maximum PHE signal in the field axis for the ring sensors. However, the peak-to-peak PHE voltages ($\Delta V_{PHE}$) were increased. Thus, the slope of the linear zone, namely the sensitivity, was increased. To further increase the PHE sensitivity, the number of the rings can be increased or the diamond shaped PHE sensors can be fabricated[10].

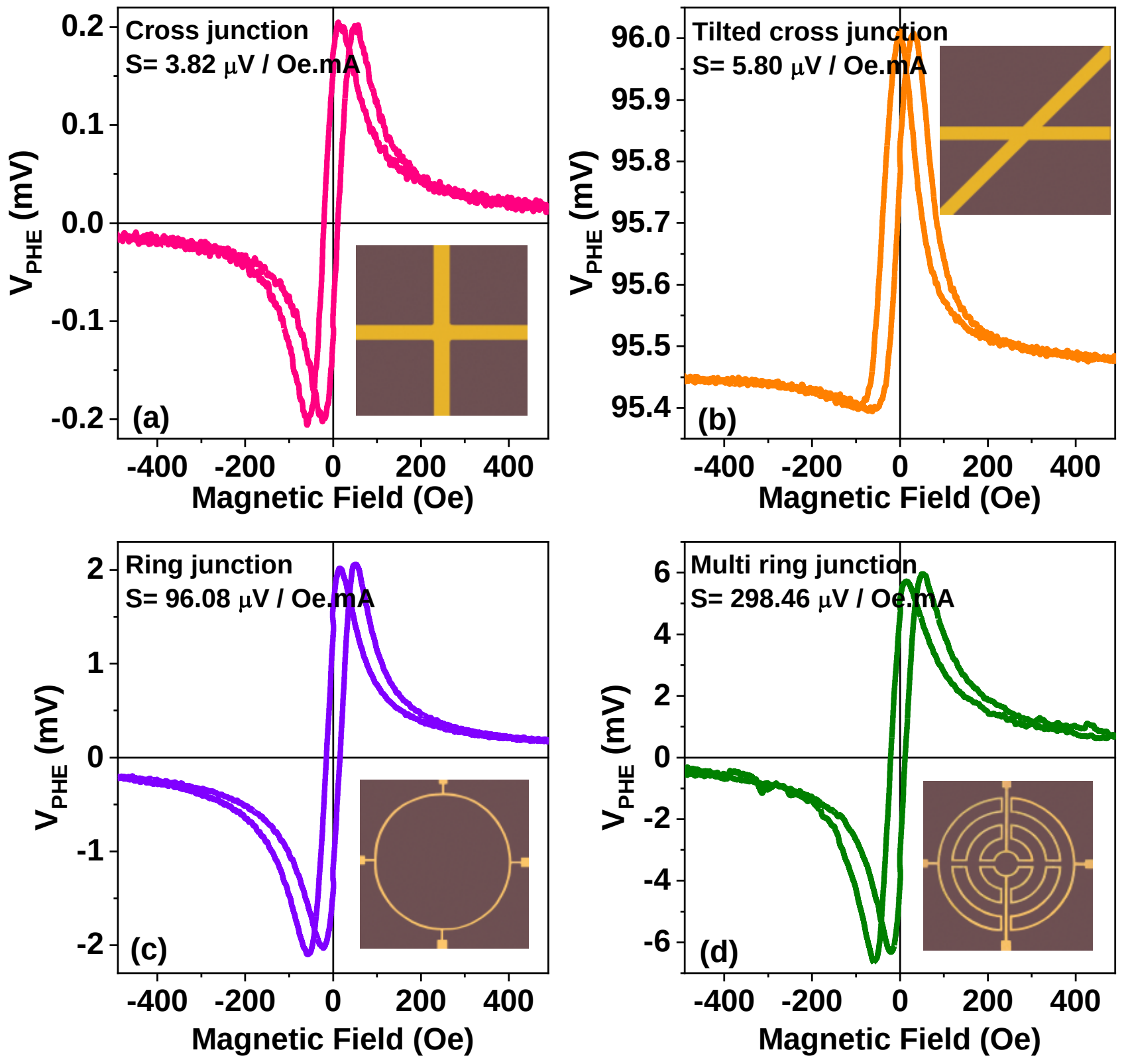

Figure 5. PHE sensor signals for the cross (a), tilted-cross (b), one-ring (c) and five-ring (d) junctions. All sensors were simultaneously grown on a Si substrate with a trilayer structure of NiFe(10)/Pt(0.8)/IrMn(8)/Pt(3) (nm).

In addition, the stability of the sensor states under different magnetic fields and the reproducibility of the sensor signal are crucial for real applications of the PHE sensors. With this motivation, we have carried out time-profile experiments presented in Figure 6. Firstly, the sensor output voltage was recorded for one minute under zero magnetic field. Then, a constant in-plane magnetic field was applied perpendicular to the current line for one minute. This process was repeated three times to see whether the sensor voltage goes back to the initial offset value. Subsequently, we have increased the magnetic field from 4 Oe to 20 Oe with a step of 4 Oe and repeated the same measurement cycle. The real-time profiles of the PHE sensors indicate that the sensor output voltages completely turned to their initial values, except with a small offset in one-ring sensor presented by blue colour. We have also observed very stable sensor states at different magnetic fields.

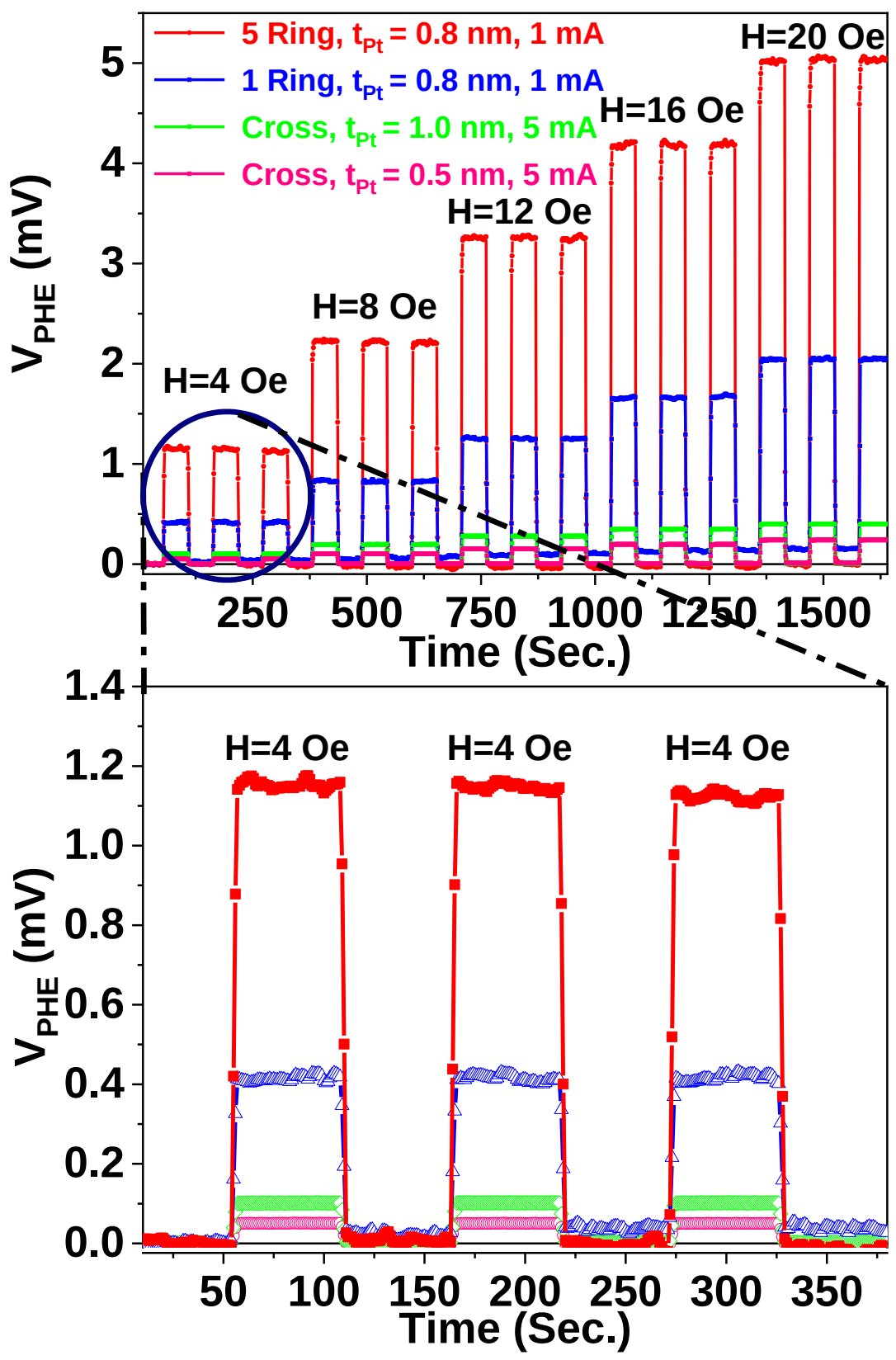

Figure 6. Real-time profile measurements of different PHE sensors recorded at 300 K. The figure at the bottom shows a magnified version of the real-time signals for a magnetic field of 4 Oe.

## 4. Discussion

The origin of the hysteretic-like behaviour observed in the PHE signals was attributed to an incoherent magnetization rotation during the positive and negative field sweepings[17,27]. In order to further understand this behaviour, we have performed PHE, perpendicular AMR and L-MOKE measurements at the hard axis geometry for a sample of 5 Å Pt inserted cross junction. Figure 7 shows that the coercive fields of the hard axis L-MOKE curve and the peak positions of the perpendicular AMR signal overlap with the hysteresis of the PHE signal. During the fabrication of the PHE sensors, the external magnetic field was zero. Thus, the growth induced anisotropy was achieved by oblique deposition. However, the data presented in Fig. 7 indicates that few portions of the exchange-biased regions slightly deviate from the main easy axis of the system during the magnetization reversal process. The observed remanent magnetization in the hard axis L-MOKE curve support this conclusion.

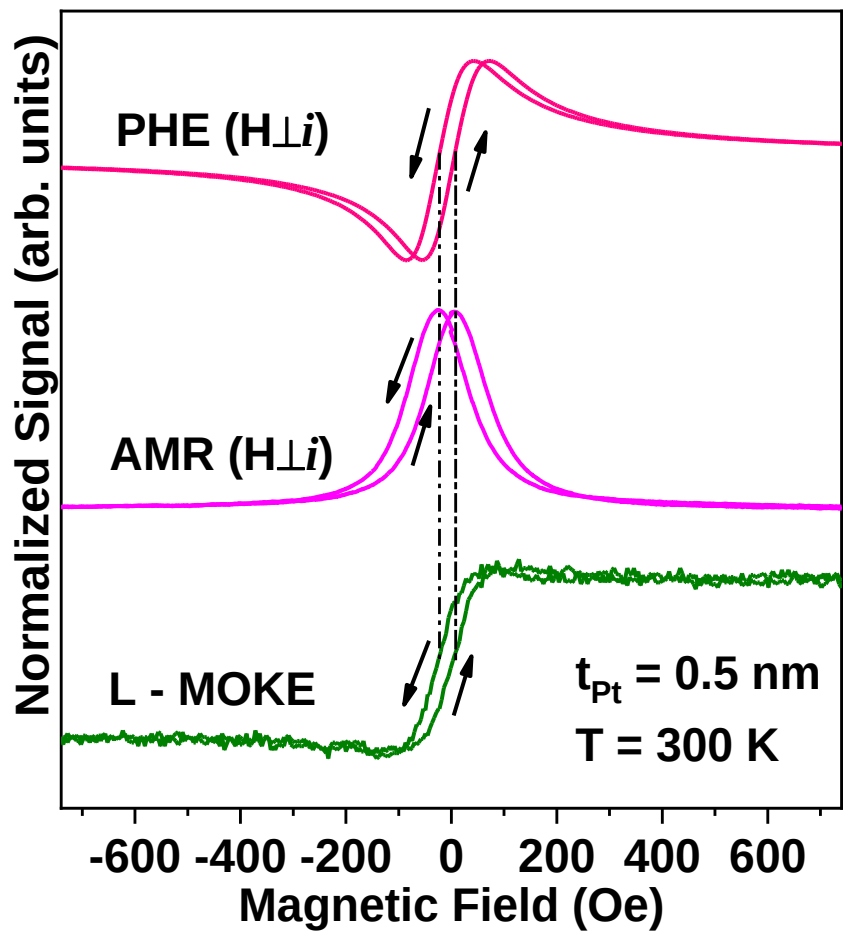

Figure 7. The PHE, perpendicular AMR and hard axis L-MOKE measurements taken from a cross junction with a Pt spacer thickness of 5 Å.

## 5. Conclusion

We have performed a systematic study of PHE sensor sensitivity in NiFe/Pt/IrMn-based trilayer structures as a function of Pt spacer thickness and sensor geometry. The PHE voltage profile measurements indicate that the sensitivity of the sensors with a cross junction architecture gradually increased up to 5.32 μV/Oe.mA by inserting thicker Pt spacers. It is worthy of note that the peak-to-peak PHE voltage ($\Delta V_{PHE}$) and AMR ratio of the sensors were increased with an oscillatory behaviour as a function of Pt thickness. Temperature-dependent experiments reveal that the PHE sensors kept working even at 390 K with a good stability, high sensitivity and high signal-to-noise ratio. In addition, we have obtained very large PHE sensitivity of 298.5 μV/Oe.mA for a five-ring sensor architecture. The real-time profile measurements indicate that the PHE sensor output voltages were reproducible and reversible at different field conditions. This makes NiFe/Pt/IrMn-based PHE sensors very promising for practical detection of low magnetic fields.

## Acknowledgements

This work was supported by TÜBİTAK (The Scientific and Technological Research Council of Turkey) through the project number 116F083.